# The role of antiphase boundaries during ion sputtering and solid phase epitaxy of Si(001)

J. C. Kim [a,b*], J. –Y. Ji [b], J. S. Kline [a], J. R. Tucker [a], and T. –C. Shen [b]

[a] Department of Electrical and Computer Engineering, University of Illinois, Urbana, IL 61801
[b] Department of Physics, Utah State University, Logan, UT 84322

---

[*] Corresponding author. Department of Physics, Utah State University, UT 84322, USA. Tel: 435-797-7070; FAX: 435-797-2492; E-mail: jkim14@staff.uiuc.edu

**Abstract**

The Si(001) surface morphology during ion sputtering at elevated temperatures and solid phase epitaxy following ion sputtering at room temperature has been investigated using scanning tunneling microscopy. Two types of antiphase boundaries form on Si(001) surfaces during ion sputtering and solid phase epitaxy. One type of antiphase boundary, the AP2 antiphase boundary, contributes to the surface roughening. AP2 antiphase boundaries are stable up to 700 $^{o}$C, and ion sputtering and solid phase epitaxy performed at 700 $^{o}$C result in atomically flat Si(001) surfaces.

The roughening of semiconductor surfaces during epitaxial growth and ion sputtering has been studied for its technological importance in semiconductor device fabrication. During epitaxial growth, adatoms attach to ascending steps at a higher rate than to descending steps or nucleate to form adatom islands depending on the substrate temperature, adatom flux and terrace width. This asymmetry in attachment kinetics of adatoms at step edges is due to the Ehrlich-Schwoebel barrier [1-4], and destabilizes the surface morphology toward the formation of mounds during multilayer epitaxial growth [5,6]. Since adatom islands nucleate independently, coalescence of two adatom islands on dimerized Si(001) [7-12] and Ge(001) [11,12] surfaces may result in the formation of antiphase boundaries. An AP1 antiphase boundary forms when two adatom islands meet at $S_A$ steps, and an AP2 antiphase boundary forms when two adatom islands meet at $S_B$ steps. Only AP2 antiphase boundaries act as preferential nucleation sites of new adatom islands during subsequent epitaxial growth [7-12].

On the other hand, the surface morphology during ion sputtering at elevated temperatures is governed by asymmetry in attachment kinetics of vacancies at step edges [13,14]; vacancies attach preferentially to descending steps. Much like during epitaxial growth, antiphase boundaries form on Si(001) surfaces during ion sputtering [15] and annealing after submonolayer ion sputtering [15,16]. However, these experiments revealed only one type of antiphase boundary, the AP2 antiphase boundary, in contrast to two formed during epitaxial growth [7-12], and the role of antiphase boundaries in roughening of crystalline semiconductor surfaces has not been investigated.

The surface morphology during solid phase epitaxy (SPE) of amorphous Si, deposited [17,18] or produced by ion sputtering [18-20] at room temperature, has been studied for its application in controlled doping and growth of Si, and activation of dopants and recrystallization following ion implantation. Antiphase boundaries form on recrystallized Si(001) [17] and Ge(001) [20]. Zandvliet and de Groot proposed that, in addition to the asymmetric kinetics of vacancies at step edges, development of antiphase boundaries may drive the Ge(001) surface roughening during ion sputtering and subsequent annealing [20]. However, since their Ge(001) surfaces were amorphized by ion sputtering at room temperature and then annealed, the role of antiphase boundaries in roughening of crystalline semiconductor surfaces was not clear in their experiments.

In this report, we present scanning tunneling microscopy (STM) data on Si(001) surface morphology during ion sputtering at elevated temperatures and SPE following ion sputtering at room temperature. We find that antiphase boundaries play an important role in surface roughening during ion sputtering by offering preferential nucleation sites of adatom islands. We also find that, due to their thermal stability below 700 $^{o}$C, antiphase boundaries may slow the relaxation of non-equilibrium Si(001) surface morphology during SPE.

Our samples are cut from either n- or p-type 0.1 $\Omega$-cm Si(001) wafers. The samples are cleaned by standard RCA procedure [21], etched by diluted HF solution, and then rinsed in dissolved oxygen free water. For the study on the Si(001) surface morphology during ion sputtering at elevated temperatures, the atomically clean and flat starting surfaces are produced by flashing the samples at ~ 1200 $^{o}$C. For the study on SPE of Si(001) surfaces, the amorphous starting surfaces are produced by 400 eV Ar ion sputtering of wet chemically cleaned samples at room temperature with the fluence of ~ 4 × $10^{17}$ ions cm$^{-2}$. Ion sputtering is performed in a UHV chamber backfilled with Ar to 5 × $10^{-5}$ Torr. Typical ion beam current is ~ 10 µA corresponding to an estimated ion flux of ~ $3 \times 10^{13}$ ions cm$^{-2}$ s$^{-1}$. The angle of incident ions is 50° from the surface normal. The samples are heated by direct current resistive heating, and the sample temperature is measured by thermocouple and infrared optical pyrometer. After ion sputtering at elevated temperatures and SPE, the samples are allowed to cool to room temperature and then imaged by STM.

STM images of the Si(001) surfaces after ion sputtering at 350 °C are shown in Fig. 1: ~ 0.7 monolayer was removed by 400 eV Ar ions in Figs. 1(a) and 1(b), and ~ 3 monolayers were removed by 1.5 keV Ar ions in Figs. 1(c) and 1(d). Two types of antiphase boundaries are observed in Fig. 1(b): two parallel dimer rows are separated by one and a half dimer row across the AP1 antiphase boundaries, and two dimer rows are misaligned by half a dimer row across the AP2 antiphase boundaries [7-9]. (The illustration of these two types of antiphase boundaries can be found in Fig. 1(b) in Ref. 9.) Antiphase boundaries form when two independently nucleated vacancy islands with different dimer row registry coalesce. An AP1 antiphase boundary forms when two vacancy islands meet with two $S_B$ steps between them, much like an AP1 antiphase boundary forms when two adatom islands meet at $S_A$ steps during epitaxial growth [7-9]; dimer rows are parallel to the $S_A$ steps and perpendicular to the $S_B$ steps [22]. An AP2 antiphase boundary forms when two vacancy islands meet with two $S_A$ steps between them; the islands decorating AP2 antiphase boundaries have two $S_A$ steps across the AP2 antiphase boundaries, see Fig. 1(b). Bedrossian observed AP2 antiphase boundaries formed on Si(001) surfaces during ion sputtering and annealing [15,16], but he did not observe AP1 antiphase boundaries.

Along the AP2 antiphase boundaries, dangling bond density is twice as large as the rest 2 × 1 surface [17], and therefore AP2 antiphase boundaries become preferential nucleation sites of adatom islands. During ion sputtering at elevated temperatures, adatoms are either directly created by sputtering removal of surface atoms or detached from step edges. After multilayer erosion by 1.5 keV Ar ions as shown in Figs. 1(c) and 1(d), both types of antiphase boundaries are still observed and the Si(001) surface morphology resembles that observed during multilayer epitaxial growth of Si by UHV CVD [9]. This resemblance suggests that AP2 antiphase boundaries play an important role in surface roughening during ion sputtering by offering preferential nucleation sites of adatom islands.

Sputtering by 400 eV Ar ions for 20 min at 700 $^{o}$C reveals an atomically flat Si(001) surface without adatom islands on wide terraces (data not shown) by step retraction. This result is in good agreement with Bedrossian's observation on the disappearance of adatom islands decorating AP2 antiphase boundaries above 700 $^{o}$C [15,16].

An STM image of the Si(001) surface after removing ~ 450 monolayers by 400 eV Ar ions at 500 $^{o}$C is shown in Fig. 2. Surface patterns of pits and mounds are observed. This surface morphology is reminiscent of the surface morphology of Ge(001) [13] and Ge(111) [14] during ion sputtering, and this similarity in the surface morphology suggests that asymmetry in attachment kinetics of vacancies at step edges is responsible for Si(001) surface roughening during ion sputtering at elevated temperatures. As mentioned before, AP2 antiphase boundaries also play an important role in surface roughening. Asymmetric kinetics of vacancies and preferential nucleation of adatom islands at AP2 antiphase boundaries control the kinetics of vacancies and adatoms respectively, and therefore we conclude that both mechanisms collaboratively contribute to Si(001) surface roughening during ion sputtering at elevated temperatures. The effects of AP2 antiphase boundaries on surface roughening may be prominent when vacancy islands can coalesce on wide terraces, i.e. when the ion fluence is low. However, as surface patterns form and terrace width becomes narrow, vacancies are more likely to attach to step edges and AP2 antiphase boundaries are less likely to form. On the contrary, asymmetric kinetics of vacancies should play a constant role in surface roughening.

STM images of the Si(001) surface morphology after SPE are shown in Fig. 3. Randomly distributed narrow rectangular terraces form with 2 × 1 dimer reconstruction on most of the surface after SPE for 5 min at 500 $^o$C in Figs. 3(a) and 3(b), and the surface is relatively rough. (Fuzzy step edges in Fig. 3(b) are due to STM tip effects, not due to incomplete recrystallization.) Terraces do not become more regular after SPE for 20 min at 550 $^o$C in Figs. 3(c) and 3(d), and two types of antiphase boundaries are clearly observed. These results on recrystallization of Si(001) surfaces are consistent with amorphous-to-crystalline transition temperature 500-600 $^o$C reported in Ref. 23.

The Si(001) surface shown in Fig. 3(c) is not much smoother than that shown in Fig. 3(a). Here, we derive an estimated time needed for relaxation of the surface morphology shown in Fig. 3(c). In 1-D, non-conserved, step-mobility limited model, which describes well the relaxation of non-equilibrium Ge(001) surface morphology [24,25], the amplitude of the surface roughness decays as $z_o(1 + t/\tau)^{-1/2}$, where $z_o$ is the initial amplitude of the surface roughness, $t$ is the time, and $\tau$ is the relaxation time expressed as

$$\tau^{-1} \cong \eta \frac{\pi^2 (kT)^2}{\beta} \left(\frac{2\pi}{L}\right)^4 \left(\frac{z_o}{a}\right)^2, \qquad (1)$$

where $\eta$ is the step mobility, $\beta$ is the step stiffness, $L$ is the characteristic in-plane length scale of the surface morphology, and $a$ is the step height. Using $\eta kT \approx 10^2$ nm$^3$ s$^{-1}$ at 550 °C, $\beta \approx 0.03$ eV nm$^{-1}$ for Si [25], $L \approx 70$ nm and $z_o \approx 5a$ from Fig. 3(c), we find $\tau \approx 0.3$ s. However, this short relaxation time cannot explain the relatively rough surface morphology in Fig. 3(c), since the surface should flatten rapidly once fully recrystallized probably after less than 5 min of SPE at 550 °C, considering crystalline surface shown in Fig. 3(a). We do not know why the Si(001) surface morphology remains relatively rough in Fig. 3(c), but the thermal stability of AP2 antiphase boundaries below 700 °C may be responsible for slow relaxation of the surface morphology.

STM images of the Si(001) surface morphology after SPE at 700 °C are shown in Fig. 4. AP1 and AP2 antiphase boundaries are observed in Figs. 4(a) and 4(b), but not in Fig. 4(c); antiphase boundaries are rarely observed after 5 min of SPE at 700 °C. This result implies that annealing time as well as substrate temperature is also an important parameter for the annihilation of antiphase boundaries. We also observe vacancy line defects, which are perpendicular to dimer rows and are probably formed by ordering of vacancies [26,27] in Fig. 4.

The role of antiphase boundaries during ion sputtering and solid phase epitaxy has been investigated. We conclude that preferential nucleation of adatom islands at AP2 antiphase boundaries as well as asymmetric kinetics of vacancies drives Si(001) surface roughening during ion sputtering at elevated temperatures. Also, thermal stability of AP2 antiphase boundaries below 700 $^{o}$C may slow the relaxation of Si(001) surface morphology during SPE.

This work was supported by NSF under Grant No. 9875129 (TCS), ARDA/ARO under Grant No. DAAD19-00-R-0007, and DARPA QuIST under Grant No. DAAD19-01-1-0324.

**Figure captions**

Fig. 1. STM images of the Si(001) surfaces after Ar ion sputtering at 350 °C. The ion energy and thickness removed are: (a) and (b) 400 eV, ~ 0.7 monolayer, (c) and (d) 1.5 keV, ~ 3 monolayers.

Fig. 2. STM image of the Si(001) surface after 400 eV Ar ion sputtering removal of ~ 450 monolayers at 500 °C.

Fig. 3. STM images of the Si(001) surfaces after solid phase epitaxy. The substrate temperature and annealing time are: (a) and (b) 500 °C, 5 min, (c) and (d) 550 °C, 20 min.

Fig. 4. STM images of the Si(001) surfaces after solid phase epitaxy at 700 °C. The annealing time is: (a) 5 s, (b) 3.5 min, (c) 5 min.

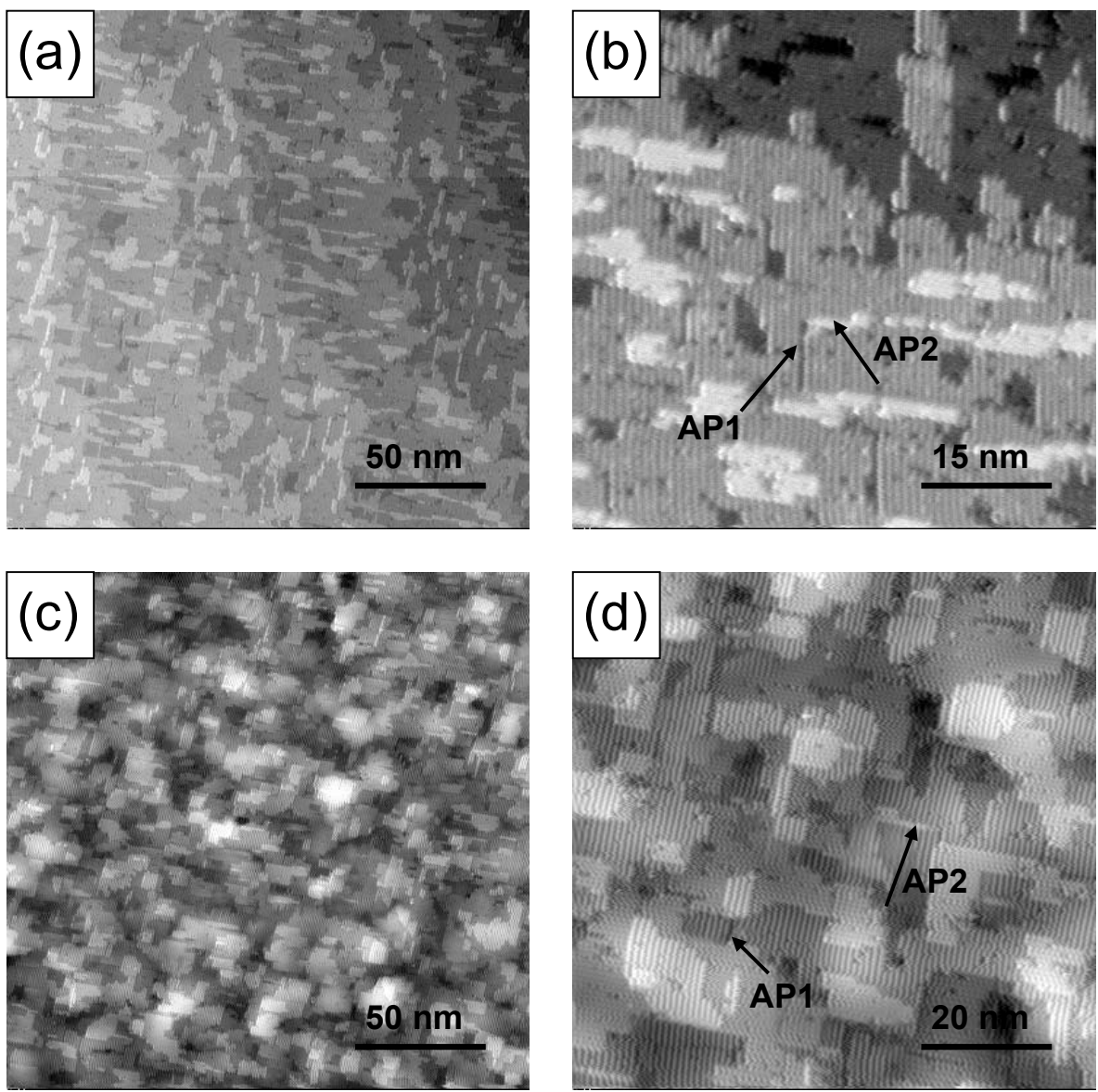

Fig. 1

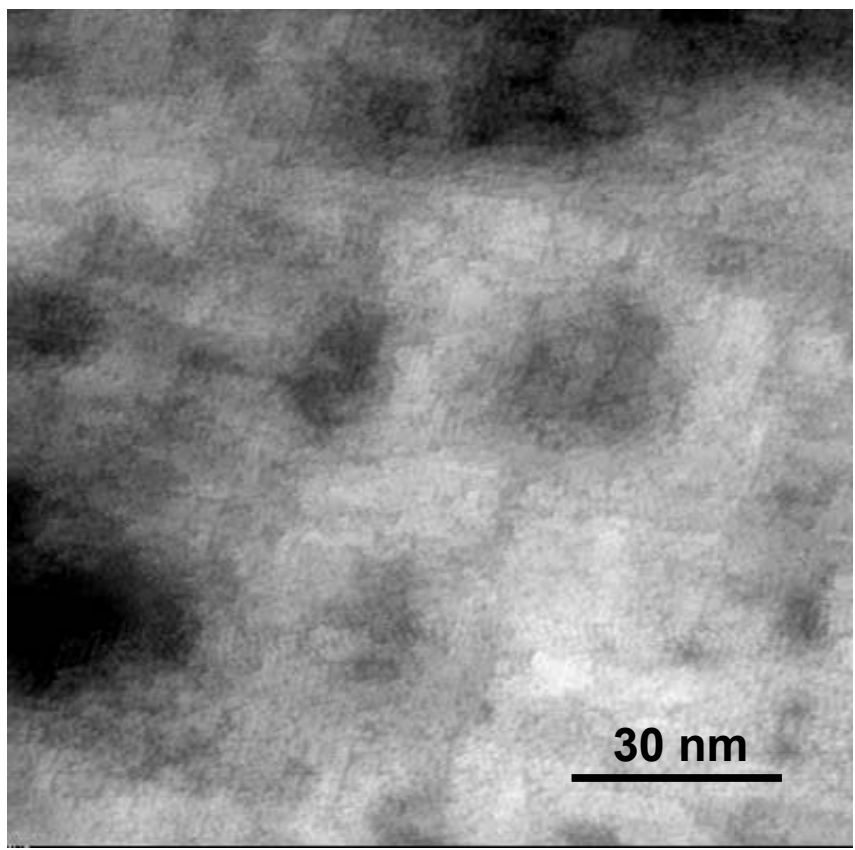

Fig. 2

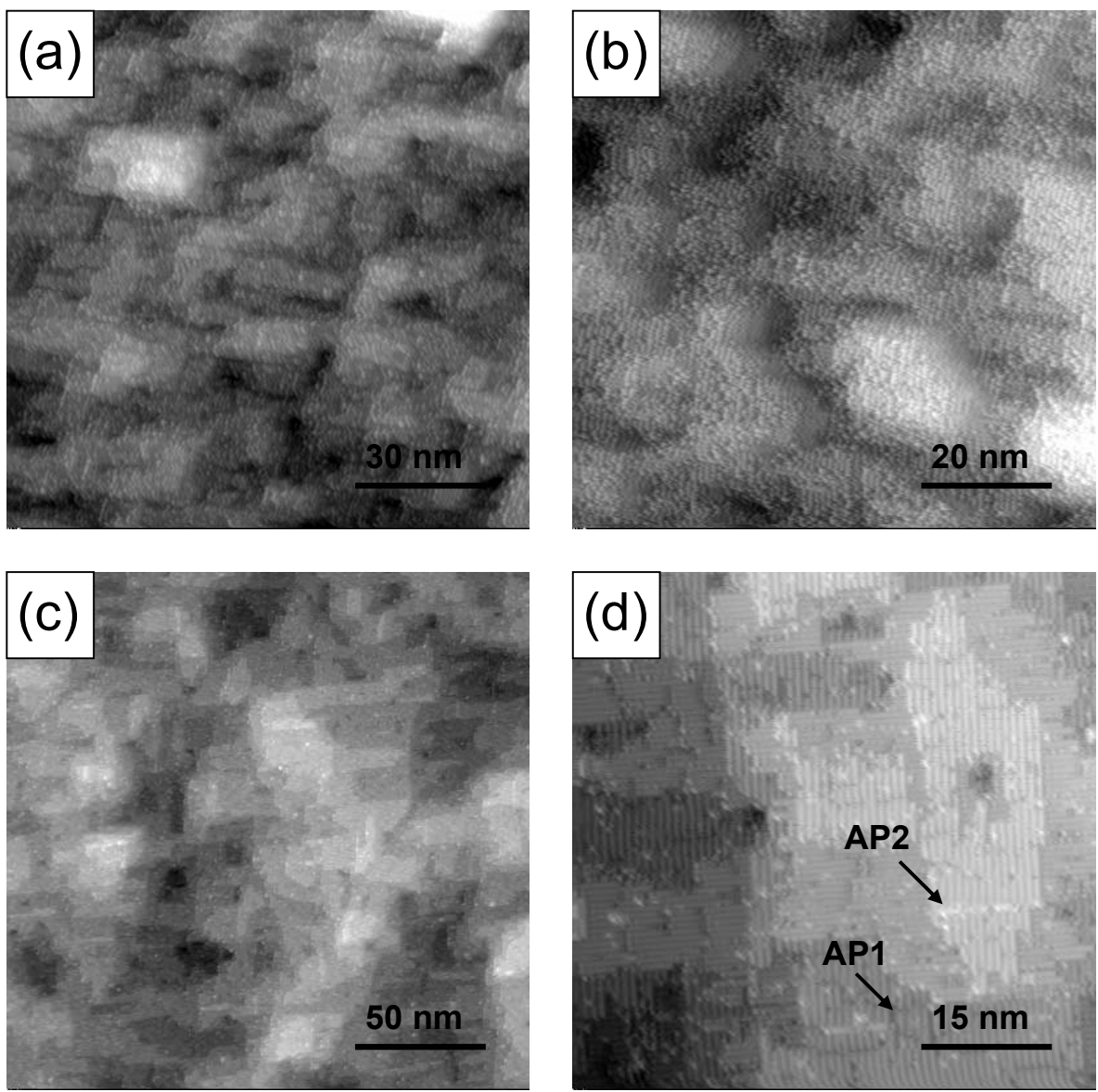

Fig. 3

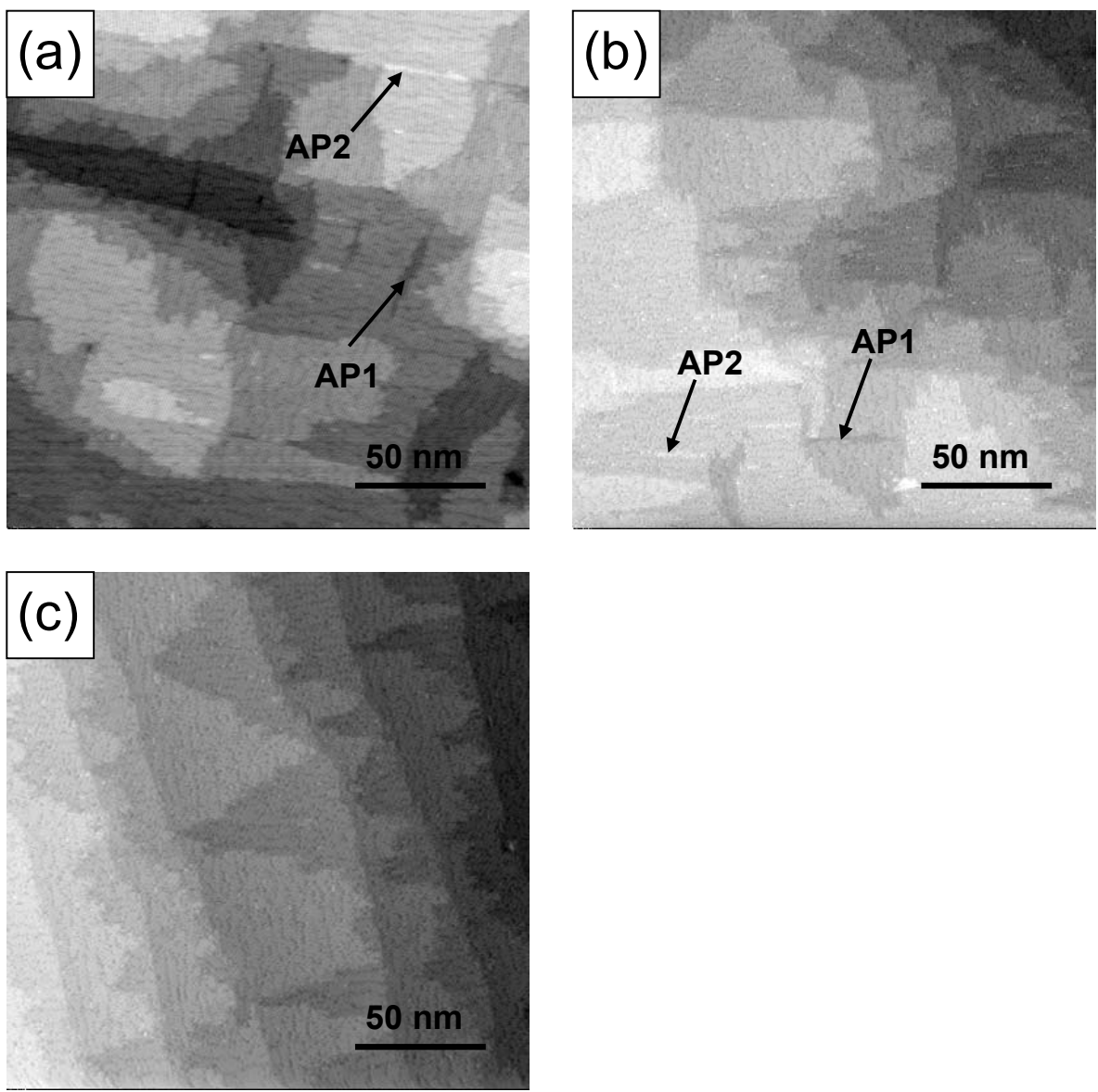

Fig. 4